\documentclass[prper,aps,twocolumn,groupedaddress,floats,showpacs,final,
superscriptaddres]{revtex4-1}
\usepackage[latin3]{inputenc}
\usepackage[makeroom]{cancel}
\usepackage{graphicx}
\usepackage{amsmath}
\usepackage{amsfonts}
\usepackage{amssymb}
\usepackage{color}
\usepackage[left=2cm,right=2cm,top=2cm,bottom=2cm]{geometry}

\usepackage{dcolumn}
\usepackage{bm}
\usepackage{simplewick}
\usepackage{array}
\usepackage{appendix}

\RequirePackage[
   hyperindex,colorlinks,bookmarksnumbered,
   plainpages=true,pdfstartview=FitH]{hyperref}
\hypersetup{linkcolor=blue,urlcolor=blue,citecolor=blue}
\usepackage{hyperref}

\definecolor{purple}{rgb}{0.5,0,0.6}

\usepackage{ulem}
\renewcommand{\emph}[1]{\textit{#1}}
\definecolor{darkblue}{rgb}{0,0,0.5}
\definecolor{darkgreen}{rgb}{0,0.5,0}
\definecolor{darkred}{rgb}{.7,0,0}
\definecolor{purple}{rgb}{0.5,0,0.6}
\definecolor{orange}{rgb}{1,0.5,0}
\definecolor{grey}{rgb}{.6,.6,.6}
\definecolor{lightpink}{rgb}{1,0.7,0.75}
\definecolor{pink}{rgb}{1,0.4,0.58}
\definecolor{deeppink}{rgb}{1,0.08,0.58}

\renewcommand{\emph}[1]{\textit{#1}}

\begin{document}

\date{\today}
\title{Landau-Zener Transitions and Rabi Oscillations in a Cooper-Pair Box:\\
Beyond Two-Level Models}

\author{A. V. Parafilo}
\author{M. N. Kiselev}
\affiliation{The Abdus Salam International Centre for Theoretical
Physics, Strada Costiera 11, I-34151 Trieste, Italy}


\date{\today}

\begin{abstract}
We investigate quantum interference effects in a superconducting Cooper-pair box by taking into account the possibility of tunneling processes involving one and two 
Cooper pairs. The quantum dynamics is analysed in a framework of three-level model.
We compute Landau-Zener probabilities for a linear sweep of the gate charge and
investigate Rabi oscillations in a periodically driven three-level system 
under in- and off- resonance conditions.
It was shown that the Landau-Zener probabilities reveal two different patterns: "step" and "beats"-like behaviours associated with the quantum interference effects.  
Control on these two regimes is provided by change of 
the ratio between two characteristic time scales of the problem. 
We demonstrate through the analysis of a periodically driven three-level system, that if a direct transition between certain pairs of levels is allowed and fine-tuned to a resonance, the problem is mapped to the two-level Rabi model. If the transition between pair of levels is forbidden, the off-resonance Rabi oscillations involving second order in tunneling processes are predicted. This effect can be observed by measuring a population difference slowly varying in time between the states of the Cooper-pair box characterised by the same parity.
\end{abstract}


\maketitle
\section{Introduction}

Time-evolution of a quantum mechanical system characterized by a discrete energy spectrum allows energy level crossings in certain situations. When two levels cross under a modulation  of some external parameter (e.g. magnetic and electric fields etc.) varying in time, the level crossings may or may not convert to avoided crossings. If the symmetry of the quantum mechanical problem
permits a cross-talk between the levels, the levels start to repel each other. 

The simplest problem where the avoided level crossing arises
is the Landau-Zener (LZ) problem \cite{landau},\cite{zener}, see \cite{nakamurabook}. The LZ Hamiltonian \cite{landau},\cite{zener} addressing a time evolution of a two-level system (TLS)
has been suggested in 1932 to describe the crossing of molecular terms aiming to construct a qualitative theory of a pre-dissociation. 
The same year, Majorana considered
a completely different problem which nevertheless falls 
into the same universality class. Namely,
Majorana \cite{majorana} investigated the behaviour of atoms subject to the time-dependent
magnetic field. The pioneering work of Majorana \cite{majorana} has anticipated the 
revolution in quantum manipulation of few-level artificially prepared
quantum mechanical systems well before the era of quantum information 
processing began (see, for example \cite{nielsen}). Quantum interference is yet another important phenomenon
appearing when two levels cross several times under modulation of an external field \cite{stueckelberg}. In particular, a periodically driven two-level system is characterized by an interference pattern known as St\"uckelberg oscillations, see review
\cite{shevchenko}.

There are several realizations of TLS based on spintronics of 
quantum dot artificial atoms \cite{dots1, dots2}, quantum beats engineered with ultra-cold gases
\cite{mark}, \cite{bloch} and superconducting  devices \cite{chargequbit1}, see, e.g., reviews \cite{makhlin},\cite{shumeiko}. Among the superconducting qubits, the quantum devices built with mesoscopic Josephson junctions allow an unprecedented level of control on quantum coherence phenomena \cite{qubit},\cite{martines}.
The charge qubit based on a Cooper-pair box (CPB) has been one of the first quantum devices
to provide the evidence of quantum interference associated with Landau-Zener-St\"uckelberg-Majorana (LZSM) physics in a non-atomic system. However, the real
CPB can be considered as the TLS only under certain approximations.
The experiments of the Helsinki group \cite{hakonen1},\cite{hakonen2} have clearly demonstrated that the interference
pattern of St\"uckelberg oscillations cannot be fully explained by the 
two-level models. On one hand, the models of quantum interferometers constructed by adding few extra levels to the two-level system may provide a suitable explanation of the experimental puzzles \cite{demkovosherov}-\cite{ashhab}. On the other hand, the models describing 
multi-level interferometers contain some additional parameters which
can be used for fine-tuning quantum systems to certain resonance transitions
and therefore inspire new experiments.

In this paper we consider a three-level model for describing the quantum dynamics of the superconducting Cooper-pair box. 
The paper is organized as follows: in Section II we introduce the CPB model
and investigate quantum dynamics associated with
Landau-Zener tunneling in three-level system under a linear-in-time sweep. 
In Section III we consider a periodically driven three-level system
and discuss in- and off- resonance Rabi oscillations. Concluding remarks are given in 
the Section IV.

\section{Landau-Zener tunneling in a Cooper Pair Box}

We consider a superconducting Cooper-pair box  -- a small superconducting island coupled both to a massive electrode via resistive Josephson junction and to a electrostatic gate via capacitance. The Hamiltonian describing this system is given by:
\begin{equation}\label{Hamil1}
H_{CPB}=E_C(\hat n-n_g)^2+E_J\cos\hat\phi.
\end{equation}
The first term in $H_{CPB}$ represents the charge states:
here $E_C{=}(2e)^2/2C$ is a charging energy of superconducting island ($C$ is its capacitance), the operator $\hat n$ accounts for the number of Cooper pairs, dimensionless gate charge $n_g{=}{-}C_gV_g/2e$ is the external parameter controlling the number of the Cooper pairs on the island via the gate voltage $V_g$. The second term in the Hamiltonian (\ref{Hamil1}) describes Josephson tunneling. Here $E_J$ is the Josephson energy and $\hat\phi$ is the phase operator canonically conjugated to 
$\hat n$: $\hat n{=}{-}i\partial/\partial \hat \phi$ (here we adopt the system of units $\hbar{=}1$). We assume that the value of a superconducting gap $\Delta_S$ of the island is larger compared to the charging energy $E_C$  ($\Delta_S{\gg} E_C$), which allows us to ignore tunneling of the odd number of charges to the island. In this paper, we investigate the charge regime $E_J{\ll} E_C$, when superconducting CPB operates as an elementary charge qubit \cite{makhlin}, \cite{shumeiko}. 
If the Josephson energy is negligibly small, $E_J{\to} 0$, a fixed number of the Cooper pairs is trapped on the island, while the ground state energy depends periodically
on the gate voltage $V_g$. Besides, there are special values of the gate voltage,
namely, $n_g(V_g){=}N{\pm} 1/2$, at which $N$ and $N{\pm} 1$ charge states become degenerate. Inclusion of the finite Josephson energy lifts the degeneracy and allows us
to approximate the CPB at low energies by a two-level system model.

In this paper we go beyond the TLS model by taking into account an additional degeneracy between $n$ and $n{+}2$ charge states occurring under condition $n_g(V_g){=}n$. The minimal model describing this case accounts for three charge states only, namely, $\{n_1,n_2,n_3\}{\equiv}\{N{-}1,N, N{+}1\}$ Cooper pairs, see Fig.1. In the regime $E_C{\gg} E_J$ the Hamiltonian is written 
in the basis formed by the charge states, parametrized by the number of Cooper pairs on the island.  The matrix form of the Hamiltonian in this basis is given by:
\begin{eqnarray}\label{Hamil2}
H=\left(
\begin{array}{ccc}
E_C(n_g-n_1)^2 & \Delta & \Sigma\\
\Delta& E_C(n_g-n_2)^2 & \Delta\\
\Sigma & \Delta & E_C(n_g-n_3)^2
\end{array}\right),\nonumber\\
\end{eqnarray}
where $\Delta{\equiv} E_J$ and $\Sigma$ are the amplitudes for tunneling on the island of one and two Cooper pairs respectively.
We start our analysis of the quantum dynamics by considering the case, when the gate voltage is swept linearly in time: $n_g(t){=}N{+}\alpha t$. In order to get simple analytical results we first restrict our analysis by imposing $\Sigma{=}0$ condition
(absence of direct tunneling of two Cooper pairs). In this case, it is easy to solve the time-dependent Schr\"odinger equation $i\dot \psi =H\cdot \psi$ with Hamiltonian (\ref{Hamil2}) by using so-called Kayanuma's method \cite{kayanuma}. The idea behind the Kayanuma's ansatz is to exclude all diagonal elements in Eq.(\ref{Hamil2}) by performing a transformation with 
a diagonal operator

\begin{eqnarray}\label{oper}
\hat U=e^{-i\theta_t}\left(
\begin{array}{ccc}
e^{-iE_C\left(\alpha t^2+t\right)} & 0 & 0\\
0 & 1 & 0 \\
0 & 0 & e^{-iE_C\left(-\alpha t^2+t\right)} 
\end{array}\right),\nonumber\\
\end{eqnarray}
where $\theta_t{=}E_C \alpha^2 t^3/3$.
\begin{figure}
\begin{center}
 \includegraphics[width=80mm]{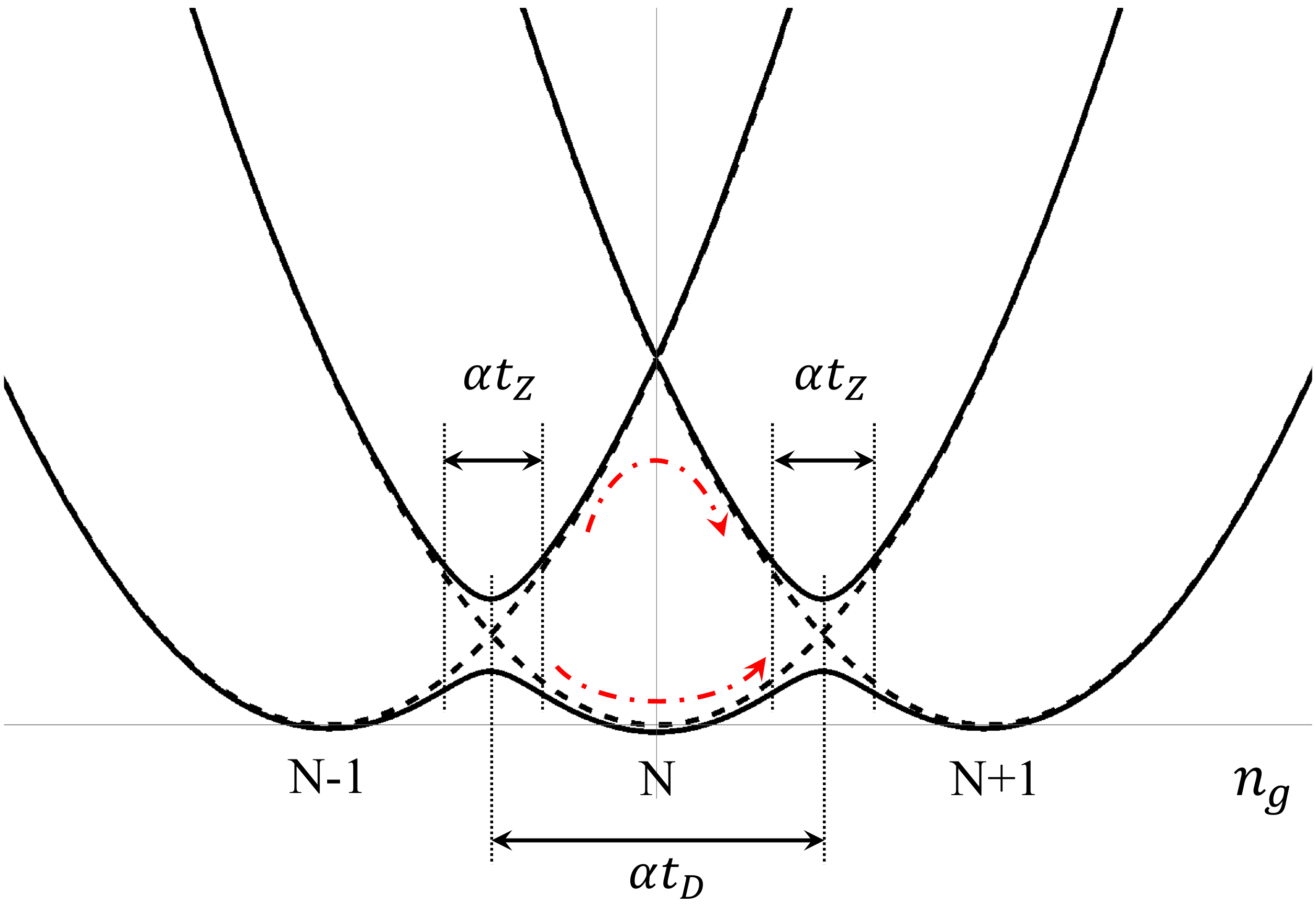}  
\caption{(Color online) The energy diagram for the superconducting Cooper-pair box model. Dashed lines denote
the charging energy given by the diagonal term of the Eq. (\ref{Hamil2})
as a function of the dimensionless gate voltage, $n_g(V_g)$ (diabatic basis
for the Landau-Zener problem). Solid lines show the adiabatic basis obtained by
diagonalization of the Eq. (\ref{Hamil2}) for a particular case $\Sigma{=}0$. Dash-dotted red curves form a closed loop and denote adiabatic and non-adiabatic paths
resulting in quantum interference.}
\end{center}
\label{fig1}
\end{figure}
Transforming the wave function $ \tilde\psi(t){=}\hat U\psi(t){=}\sum_{i=1}^3C_i(t)\vert i\rangle$, where the states $\vert i\rangle$ form the compact basis of diabatic states of Eq.(\ref{Hamil2}), we re-write the non-stationary Schr\"odinger equation describing the time-evolution of the three-level system in terms of the system of three linear differential equations:
\begin{eqnarray}\label{schr}
&& i\dot C_1(t)= \Delta e^{iE_C\left(\alpha t^2+t\right)} C_2(t), \nonumber \\
&& i\dot C_2(t)= \Delta e^{-iE_C\left(\alpha t^2+t\right)} C_1(t)+  \Delta e^{-iE_C\left(-\alpha t^2+t\right)}C_3(t), \nonumber \\
&&i\dot C_3(t)=\Delta e^{iE_C\left(-\alpha t^2+t\right)} C_2(t).
\end{eqnarray}
To find a solution of the system of coupled linear differential equations, it is
convenient to rewrite it in the form of linear integral Volterra equations.
For example, it is straightforward to transform the equation for $C_2(t)$ to a 
self-contained integral form by excluding $C_1(t)$ and $C_3(t)$
with the help of the first an the third equations in (\ref{schr}):
\begin{eqnarray}\label{amp}
&&C_2(t)=-\Delta^2\int_{-\infty}^t dt_1 \int_{-\infty}^{t_1}dt_2 C_2(t_2)\times\nonumber\\
&&\times\left\{\exp\left[-iE_C (t_1^{+})^2+iE_C (t_2^{+})^2\right]+\right.\nonumber\\
&&+\left.\exp\left[iE_C (t_1^{-})^2-iE_C (t_2^{-})^2\right] \right\},
\end{eqnarray}
where $ t^{\pm}=\sqrt{\alpha}t\pm1/(2\sqrt{\alpha})$. We assume that 
the initial condition for Eqs.(\ref{schr}) is given by the $N$ - quasiparticle
Cooper pairs state 
characterized by the occupancy: $C_2(-\infty){=}1$ and $C_{1}(-\infty){=}C_{3}(-\infty){=}0$.

The integral equation (\ref{amp}) is solved by the iterations. This procedure is legitimate in the non-adiabatic approximation under condition $\delta {=}\Delta^2/(\alpha E_C ){\ll} 1$. By exponentiating the result of the first iteration we obtain the probability $P_2{=}|C_2|^2$ to find the system in the $N$-charge state at $t{\rightarrow }\infty$:
\begin{eqnarray}\label{probability}
P_{2}(t)\approx\exp\left(-\frac{\pi}{2}\frac{\Delta^2}{\alpha E_C}\left[F\left(\tilde t^{+}\right)+F\left(\tilde t^-\right)\right]\right),
\end{eqnarray}
where the function
\begin{eqnarray}\label{def}
F(z)=\left[\left(\frac{1}{2}+C(z)\right)^2+\left(\frac{1}{2}+S(z)\right)^2\right]
\end{eqnarray}
is expressed in terms of the Fresnel integrals
\begin{equation}\label{fresnel}
S(z)=\sqrt{\frac{2}{\pi}}\int_0^zdt\sin t^2 \;, \quad C(z)=\sqrt{\frac{2}{\pi}}\int_0^zdt\cos t^2.
\end{equation} 
In Eq.(\ref{probability}) we denote $\tilde t^{\pm}{=}\sqrt{2E_C/\pi}[\sqrt{\alpha}t{\pm} 1/(2\sqrt{\alpha})]$. 
We plot on Fig.~2 the probability $P_2$ obtained by analytic solution of the Eq.(\ref{probability})  for two different sets of parameters (see details in the figure caption): an orange curve represents the solution with $\Sigma{=}0$, while a black curve corresponds to 
the solution with $\Sigma{\neq}0$. The step-like behaviour characteristic for the orange curve is originating from an interplay between two time scales of the LZ problem \cite{gefen1},\cite{gefen2}: i) a Zener time $t_Z{\sim} (\alpha E_C)^{-1/2}$ associated with the "individual" Landau-Zener transitions at corresponding avoided crossings (we consider $t_Z$
for the non-adiabatic LZ transition \cite{gefen1});  ii) a dwell time $t_D{\sim} \alpha^{-1}$ related to the time interval between two consequent crossings 
(see Fig.1 and Fig 2.). Two different regimes correspond to two opposite limiting cases: (i) two Landau-Zener transitions can be considered as two consequent (independent) avoided crossings if $t_Z{<}t_D$ (see the upper panel in Fig.2), and (ii) 
two transitions can not be separated in time if $t_Z{>}t_D$ and the interference from the nearest avoided crossings must be taken into account (see the lower panel in Fig.2). This interference results in a pronounced super-structure in the time 
evolution of the probability $P_2(t)$. Emergence of the two energy scales $E_1$ and $E_2$ with $E_1{-}E_2 {\sim }E_C$ leads to the "beats" pattern characterized by  the period 
$t_{beats}{\sim }E_C^{-1}$. 
It is convenient to consider a "triangle" formed by three parabolas (see Fig.~1)
as an Mach-Zehnder interferometer. Each avoided crossing point is equivalent 
to a "mirror" characterized by a transparency determined by LZ probability.
The left avoided crossing therefore splits the state into two parts (red dash-dotted lines representing adiabatic and non-adiabatic paths in Fig.~1), while the right crossing can either play a role of yet another splitter (if $\Sigma{=}0$) or
detect an interference between transmitted (diabatic) and reflected (adiabatic) paths if $\Sigma{\neq}0$.
The "beats" super-structure is
associated with the repopulation of all three states of 
{\color{black}the Mach-Zehnder interferometer} due to almost perfect "transmission"
at $n_g{=}N$ (induced tunneling is given by the second order processes 
$\propto \Delta^2/E_C$, see \cite{kiselev1} for the details.

The interference pattern changes its character when
the "transmission" at $n_g{=}N$ associated with the tunneling of two Cooper pairs
becomes pronounced (black curves in Fig. 2). 
The "finite reflection"
at the "upper mirror" (splitter) $n_g{=}N$ leads to the probability $P_2$ deficit
(see the difference between the orange and black curves at the upper panel of Fig.2) and modifies the step pattern in the regime $t_Z{<}t_D$.
Besides, we emphasize that the probabilities to find system in $N{-}1$, $N{+}1$ states are equally distributed in the absence of $\Sigma$-terms. The reason for equipartition is due to 
equivalence of two tunneling rates at two avoided crossing points $t=\pm 1/(2\alpha)$.  This effect holds in both regimes $t_Z {\lessgtr} t_D$. Taking into account 
finite $\Sigma$ results in appearance of an asymmetry between the probabilities $P_1$ and $P_3$. Moreover, this asymmetry becomes even more pronounced in the case 
$t_Z{>}t_D$, see inserts in lower panel of the  Fig.2.

\begin{figure}
\begin{center}
 \includegraphics[width=80mm]{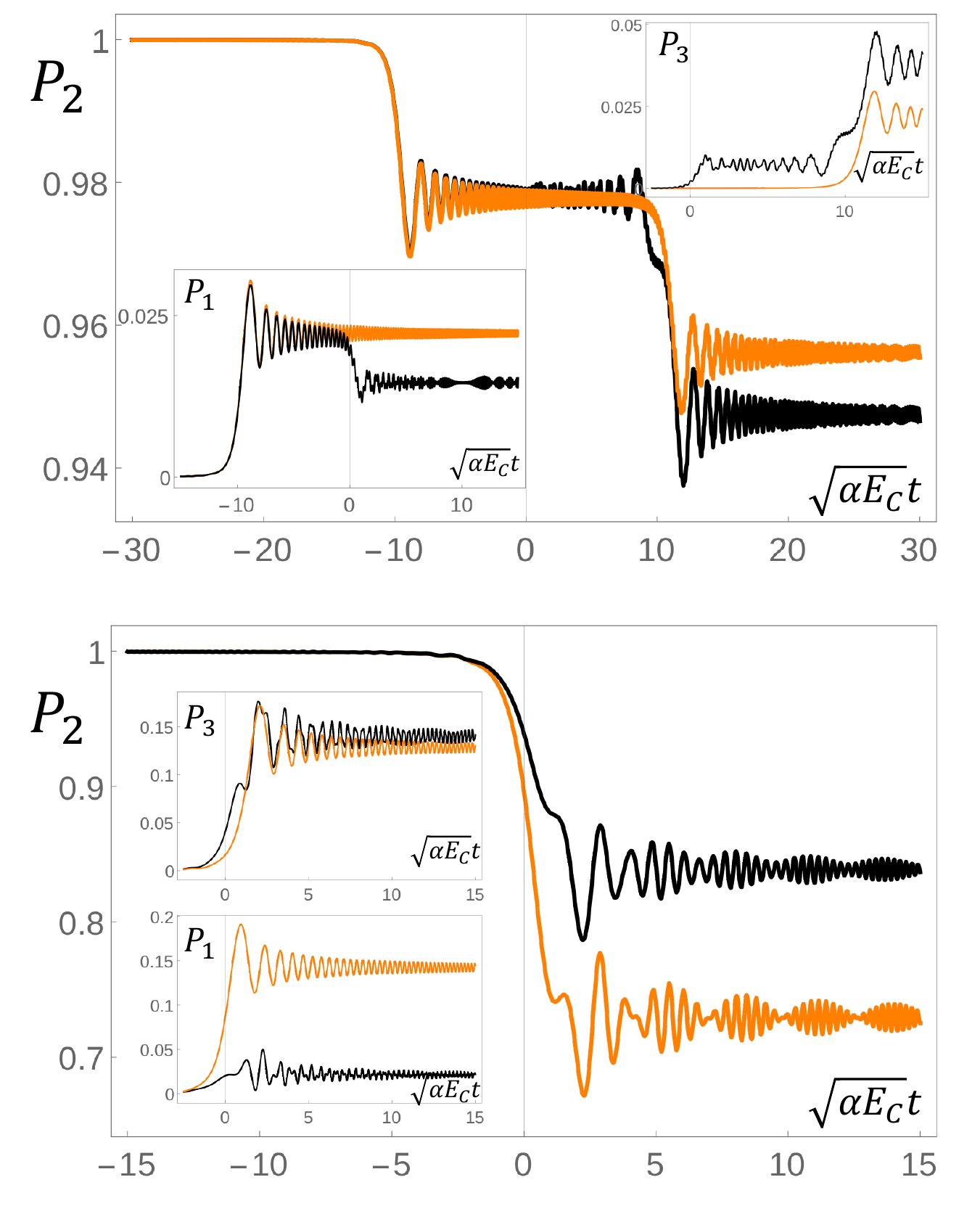}  
\caption{(Color online) 
Main frames: Time-dependent Landau-Zener probability $P_2(t){=}|C_2(t)|^2$ given by 
Eq. (\ref{schr})
as a function of a dimensionless time $\sqrt{\alpha E_C}t$ (see definitions and detailed explanations in the Section II). The inserts show the time evolution of the probabilities $P_1(t){=}|C_1(t)|^2$ and $P_3(t){=}|C_3(t)|^2$.
All curves are computed for the non-adiabatic regime $\delta{\ll}1$.
The orange curves correspond to the analytic solution Eq.~(\ref{probability}) with $\Sigma{=}0$. The black curves represent the results of numerical calculations
performed with $\Sigma{\neq}0$. Without any loss of generality we assume that the transparency at each avoided crossing point can be fine-tuned independently. We therefore do not rely upon a smallness of $\Sigma$ compared to $\Delta$.  The initial condition for all curves reads:
$C_2(-\infty){=}1$ and $C_{1}(-\infty){=}C_{3}(-\infty){=}0$.
Upper panel: $t_Z{<}t_D$,  parameters $\delta=0.0042$, $\Delta/E_C{=}0.004$ and $\Sigma/E_C{=}0.024$. Lower panel: $t_Z{>}t_D$, parameters $\delta=0.011$, $\Delta/E_C{=}0.2$ and $\Sigma/E_C{=}0.8$.}
\end{center}
\label{fig2}
\end{figure}

\section{Periodically driven CPB}

In this Section we consider a periodic modulation of the dimensionless gate charge 
\begin{eqnarray}\label{drive}
n_g(t)=N+\varepsilon_0+A\cos (\Omega_D t)
\end{eqnarray}
where $\Omega_D$ and $A$ are the frequency and amplitude of the modulation respectively and $\varepsilon_0$ is the charge offset. We investigate the cases of resonance and off-resonance drivings
and analyse Rabi oscillations \cite{rabi} in the driven three-level system.
The system is resonantly driven if the frequency of the drive $\Omega_D$
coincides with the energy difference 
between two neighbouring states (two levels).
In that case, known as a conventional Rabi problem \cite{rabi}, the probability to occupy each of two eigenstates oscillates with the frequency 
proportional to the amplitude of the drive.
When the two-level system is driven off-resonance, the oscillation frequency
$\Omega_{\rm off}{>}\Omega_R$. We show that the off-resonance driving
of the three-level system allows a strong violation of this inequality.

\begin{figure}
\begin{center}
 \includegraphics[width=80mm]{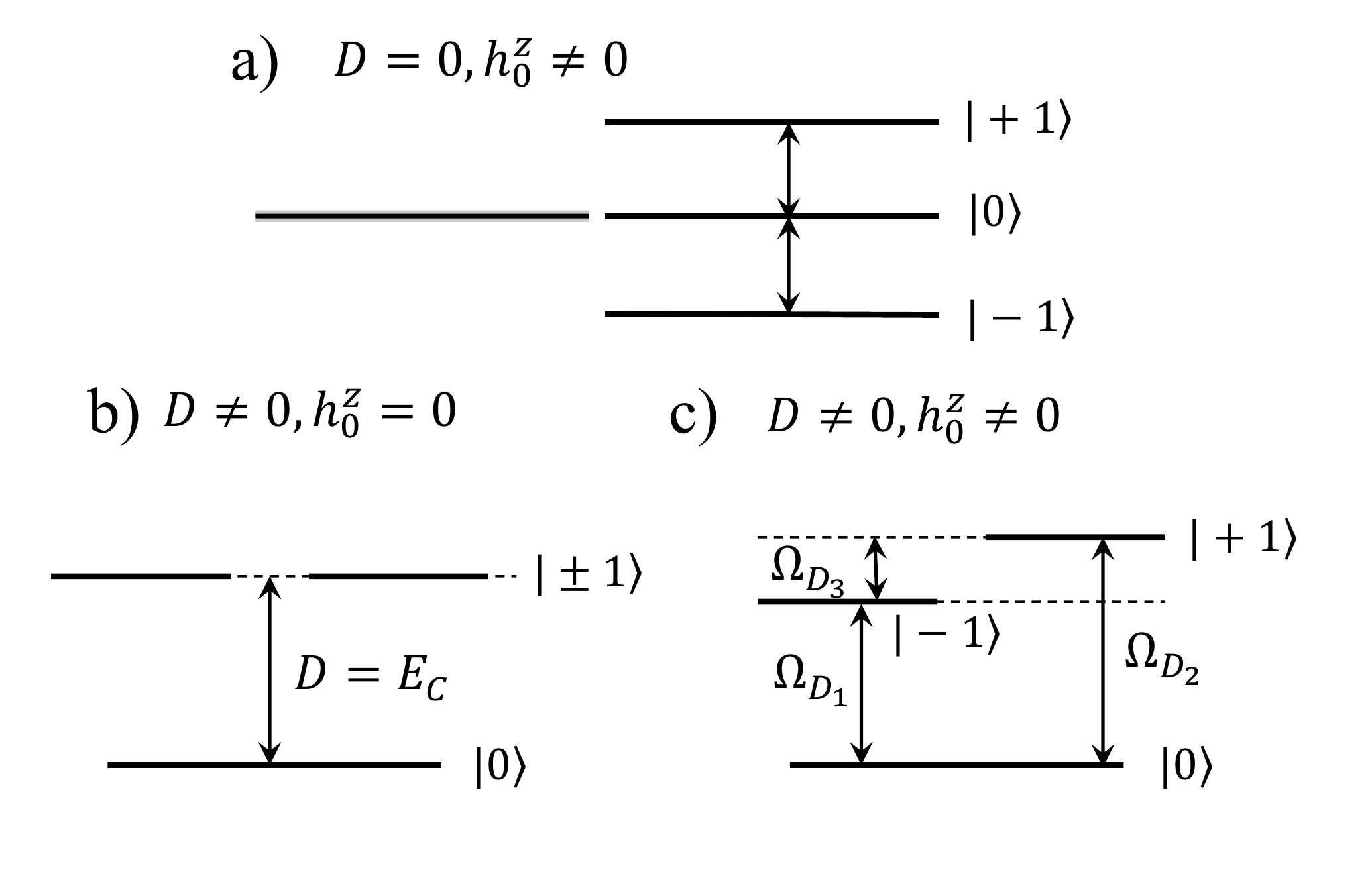}  
\caption{Energy spectrum for the $S{=}1$ model in the presence of a single-ion
anisotropy parameter $D$ (see the main text for the discussion of the mapping between
the three-level CPB models and $S{=}1$ Hamiltonians). 
(a) The single-ion anisotropy parameter 
$D{=}0$. The three-fold degeneracy of the $S{=}1$ state is lifted out by static magnetic field $h_0^z=2E_C\varepsilon_0$. Equidistant splitting of $|{\pm} 1\rangle$ states is described by a linear Zeeman effect. (b) Finite  single ion anisotropy $D{\neq}0$ lifts out the degeneracy between $|0\rangle$ and $|{\pm} 1\rangle$ states.
The states $|{\pm} 1\rangle$ still remain degenerate in the absence of magnetic field.
(c) Finite synthetic magnetic field $h_0^z{\neq} 0$ eliminates the degeneracy between $|{\pm}1\rangle$ states. When all degeneracies of the effective $S{=}1$ model are lifted out,
there exist three resonances frequencies corresponding to the transitions between three pairs of levels.  Conditions for the in- and off- resonance transitions are discussed in the Section III.}
\end{center}
\label{fig3}
\end{figure}

\subsection{Mapping three-level systems to $S{=}1$ models}
To analyse the quantum dynamics of a multi-level  CPB, it is convenient
to use an equivalent language of spin-$S$ states representing
$2S{+}1$ - levels model.
In particular, the diagonal part of the Hamiltonian describing three-level $S{=}1$
system can always be represented in terms of a linear (dipole moment) and quadratic
(quadrupole moment) combinations of $\hat S^z$. The transitions between the 
eigenstates of $\hat S^z$ operator
are accounted by linear terms in $\hat S^x$, $\hat S^y$ operators and also corresponding
bi-linear combinations (quadrupole moments).
Rewriting the Hamiltonian
(\ref{Hamil2}) in the basis of linear and bi-linear spin $S{=}1$ operators
results in the following spin Hamiltonian:
\begin{eqnarray}\label{eaxis}
{\cal H}=H-H_0(t)=\Delta \hat S^x + h^z(t) \hat S^z + D (\hat S^z)^2
\end{eqnarray}
where 
$h^z(t){=}2E_C\varepsilon_0{+}2AE_C\cos(\Omega_D t)$ is a synthetic time-dependent magnetic field, $D{=}E_C$ is an easy-axis anisotropy parameter
(quadrupole interaction) and $H_0(t){=}[h^z(t)]^2/(4E_C)$. Note, that the Eq. (\ref{eaxis})
describing three-level system
is not linear in terms of the $S$-operators, in contrast to the Hamiltonians
describing the quantum dynamics of the TLS. However, the Eq.
(\ref{eaxis}) as well as any three-state 
Hermitian Hamiltonians represented by $3\times 3$
matrices can be written down as  a linear form in a basis of Gell-Mann matrices
(generators of SU(3) group) \cite{kiselev1}. The linear in terms
of the $S{=}1$ operators part of the Hamiltonian (\ref{eaxis}) corresponding to
$D{=}0$ case falls into a class of SU(2) symmetry group. The transitions between
the eigenstates of $\hat S^z$ operator, $\{\vert{-}1\rangle, \vert 0\rangle, \vert{+}1\rangle \}$ (which are equivalent to $\{N{-}1$, $N$, $N{+}1\}$ charge states of the CPB model), are restricted by $\Delta S^z{=}\pm 1$ condition.
Constant (non-oscillating) magnetic field applied along $z$ direction, $h^z_0{=}2E_C\varepsilon_0$ lifts the three-fold degeneracy of the $S{=}1$ states (linear Zeeman effect).
Since the $\vert\pm\rangle$ states are equidistant from  the $\vert0\rangle$
state, the driving with $\Omega_D{=}h^z_0$ gives an access to the transitions
$\vert{-}1\rangle{\leftrightarrow}\vert0\rangle$ and $\vert0\rangle{\leftrightarrow}\vert{+}1\rangle$ (see Fig. 3(a)). 

Finite quadrupole interaction (single-ion anisotropy) $D{\neq} 0$
lifts out the degeneracy between $\vert0\rangle$ and $\vert{\pm}1 \rangle$ states
(see Fig. 3(b)). Finite synthetic magnetic field $h^z_0$ (aka charge offset)
applied along $z$-direction eliminates the degeneracy of $\vert{\pm} 1\rangle$ states.
Therefore, finite $D$ - term
explicitly breaks the $SU(2)$ symmetry and allows transitions with 
unrestricted selection rule $\Delta S^z{=}{\pm}2$.
However, the CPB model Eq. (\ref{Hamil2}) is derived under condition $E_C{=}D{\gg}\Delta$. Thus, the $SU(2)$ symmetric point is beyond the validity of the CPB model.

\subsection{Rotating Wave Approximation}
The diagonal elements of the Eq. (\ref{Hamil2}) subject to the periodic drive Eq. (\ref{drive}) explicitly depend on time. We perform the first (exact) step 
in transforming the Hamiltonian of the model by
rewriting Eq. (\ref{Hamil2}) in the new rotating frame basis by applying
a transformation:
\begin{eqnarray}\label{oper2}
&&\hat V=\exp\left(-iE_C\left[\frac{2A}{\Omega_D}\sin(\Omega_D t) \cdot\hat S^z+\right.\right.\\
&&\left.\left.\hat I\cdot\left(\varepsilon_0^2t+\frac{2\varepsilon_0A}{\Omega_D}\sin(\Omega_D t)+\frac{A^2}{2\Omega_D}\left(t+\frac{\sin (2\Omega_D t)}{2}\right)\right)\right]\right),\nonumber
\end{eqnarray}
The transformation Eq. (\ref{oper2}) results in elimination of the time-dependence from the diagonal matrix elements of the Eq. (\ref{Hamil2})  by transferring it 
to the off-diagonal elements of the 
Hamiltonian matrix. In Eq.(\ref{oper2}) $\hat I$ denotes the unit $3\times 3$ matrix.
Further simplification of the transformed Hamiltonian is achieved by rewriting the time-dependent off-diagonal elements of Hamiltonian matrix  with a help of the textbook identity for the Bessel functions:
 $\exp(ix\sin t)=\sum_m J_m(x)e^{imt}$. As a result, the new Hamiltonian
 $\tilde H{=}\hat V^{-1} H \hat V-i\hat V^{-1}\dot{\hat V}$ reads as follows:
\begin{eqnarray}\label{Hamil3}
\tilde H=\sum_{m=-\infty}^{\infty}\left(
\begin{array}{ccc}
 E_C (1+2\varepsilon_0) & \Delta_m e^{im\Omega_D t} & 0\\
\Delta_m e^{-im\Omega_D t} & 0 & \Delta_m e^{im\Omega_D t} \\
0 & \Delta_m e^{-im\Omega_D t} & E_C(1-2\varepsilon_0)
\end{array}\right),\nonumber\\
\end{eqnarray}
where $\Delta_m{=}\Delta J_m(2AE_C/\Omega_D)$. The wave functions $\varphi(t)$
written in the rotated basis are connected to the wave functions $\psi(t)$
in the original basis through the equation
$\varphi(t){=}\hat V^{-1}\cdot \psi(t)$. Note, that the 
Hamiltonian (\ref{Hamil3}) remains explicitly  time-dependent after 
the transformation Eq. (\ref{oper2}). 

The next step is to transform the Hamiltonian (\ref{Hamil3}) to a time-independent form. It can be done by applying the second transformation to yet another rotating frame.
Unfortunately, as is known, there is no simple way to eliminate exactly the time-dependence from the Eq. (\ref{Hamil3}). However, it can be done approximately using a reliable ansatz known as a {\it rotating wave approximation} (RWA).
The idea behind RWA is to consider the solution of the Schr\"odinger equation as a sum of the $k$-th harmonics: 
\begin{eqnarray}\label{schr2}
\varphi(t)=\sum_k\left(
\begin{array}{ccc}
e^{ik\Omega_D t} & 0 & 0\\
0 & 1 & 0 \\
0 & 0 & e^{-ik\Omega_D t}
\end{array}\right)\tilde\varphi_k(t),
\end{eqnarray}
For each $m$-th harmonic in Eq. (\ref{Hamil3}) there exists
corresponding $k{=}m$ term in Eq. (\ref{schr2}) such a way that the off-diagonal matrix element of the new Hamiltonian will be  
given by a sum of two terms: one is non-oscillating and another one is fast oscillating.
After neglecting the fast oscillating terms in Eq.(\ref{Hamil3}) we write the Schr\"odinger equation for $m$-th harmonic, $\tilde \varphi^{(m)}$ as follows:
\begin{eqnarray}\label{schr3}
i\dot {\tilde \varphi}^{(m)}(t)=\left(
\begin{array}{ccc}
E_C+\delta \omega_m & \Delta_m & 0\\
\Delta_m & 0 & \Delta_m  \\
0 & \Delta_m  & E_C-\delta\omega_m
\end{array}\right)\cdot \tilde \varphi^{(m)}(t),\nonumber\\
\end{eqnarray}
where $\delta\omega_m{=}m\Omega_D{+}2E_C\varepsilon_0$. While a general solution of Eq.(\ref{schr3}) is cumbersome, we consider below only some cases of a special interest.

\subsection{Resonance Rabi oscillations in CPB} 

The matrix form of the time-independent Hamiltonian (\ref{schr3}) assumes that
only two pairs of the levels, namely $(N, N{+}1)$ and $(N, N{-}1)$, can be 
fine tuned to the resonance by adjusting $\delta\omega_m$. The resonance between
$(N{-}1, N{+}1)$ states typically is not accessible due to the absence 
(smallness) of the corresponding matrix elements. Indeed, the 
probability for two Cooper pairs to tunnel in CPB is intuitively small due to
smallness of the phase space for such a process. 
Therefore, there are only {\color{black} two} resonance Rabi oscillations in the CPB model. If $\delta \omega_m{=} {+}E_C$ (which is equivalent to 
$m\Omega_D{=}E_C(1{-}2\varepsilon_0)$), the resonance condition for the transition between  $(N, N{+}1)$ is satisfied. 
This resonance condition assumes that the three-level system is considered away from the resonance $n_g{=}N{\pm}1/2$. It provides a low bound for the offset charge
$|\varepsilon_0|{<}(1/2)(1-\Delta/E_C)$.
The states $(N, N{-}1)$ stay off-resonance
being separated by a large energy offset $2E_C$. Under this condition 
the transition between $(N, N{-}1)$
can be neglected and the Hamiltonian matrix (\ref{schr3}) reduced to 
$2{\times} 2$ form \cite{carroll}. The Rabi oscillations in the TLS are  
described by the standard textbook equation \cite{rabi} (for simplicity we focus on a single-photon $m{=}1$ resonance): the resonance drive with 
$\Omega_D{=}E_C(1{-}2\varepsilon_0)$ results in oscillations with 
$\Omega_R{=}2A\cdot\Delta/(1{-}2\varepsilon_0)$ if amplitude of the drive 
$A{\ll} \Omega_D/E_C$ (to obtain the equation for $\Omega_R$ we use an
asymptotic of the Bessel function $J_1(z{\ll} 1){\approx} z$). 

If the TLS is driven near the $n_g{=}N{\pm}1/2$ resonance,
we expand the dimensionless gate charge across the resonance as follows:
\begin{eqnarray}\label{drive1}
n_g(t)=N\pm 1/2+\tilde\varepsilon_0+\tilde A\cos (\Omega_D t).
\end{eqnarray}
The resonance condition reads $\Omega_D{=}\Delta$ and $\Omega_R{\propto}\tilde A$
in accordance with the standard theory of the Rabi oscillations.

If $\delta \omega_m{=} {-}E_C$ (which is equivalent to 
$m\Omega_D{=}{-}E_C(1{+}2\varepsilon_0)$),  the resonance condition for a transition between  $(N, N{-}1)$ is satisfied and the states $(N, N{+}1)$ stay off-resonance.
Analysing corresponding Rabi oscillations in the TLS under the resonance condition $\Omega_D{=}{-}E_C(1{+}2\varepsilon_0)$ for the single-photon processes $m{=}1$ we obtain the Rabi oscillations
with a frequency $\Omega_R{=}{-}2A\cdot\Delta/(1{+}2\varepsilon_0)$. The analysis of the multi-photon resonances and periodic driving near the $n_g{=}N{\pm}1/2$ resonance
the can be performed similarly to the analysis of $(N, N{+}1)$ Rabi oscillations
considered above.

If $\Sigma{\neq}0$ direct tunneling of two Cooper pairs is allowed, 
the third Rabi resonance between 
$(N-1){\leftrightarrow}(N+1)$ 
states is possible. In that situation
the $N$ state is separated from $(N{\pm}1)$ states by the large energy gap
$E_C$ and therefore can be neglected. The resonance condition for the Rabi oscillations in the TLS reads as 
$\Omega_D{=}\Sigma$ and the Rabi frequency is proportional to the amplitude of corresponding drive.

\subsection{Off-resonance Rabi oscillations in CPB}

As we have pointed it out in the previous Subsection, the matrix 
element describing tunneling of two Cooper pairs is negligible compared to
the Josephson energy. Therefore, without loss of any generality we assume that
$\Sigma{=}0$ and 
there is no direct transition between $N{+}1$ and $N{-}1$. However, such transition arises as a second order tunneling process.
We are referring to Rabi oscillations associated with indirect 
$(N{+}1){\leftrightarrow}(N{-}1)$ transition as the off-resonance Rabi effect. The degeneracy of $N {+}1$ and $N{-}1$ levels
(in the absence of direct tunneling) is restored under condition $\delta\omega_m{=}0$
or $m\Omega_D{=}{-}2E_C\varepsilon_0$.
The solution of Eq.(\ref{schr3}) is written down in the form 
\begin{eqnarray}\label{solu}
\tilde \varphi^{(m)}(t){=}\frac{1}{2}\exp(-iE_C t/2)\cdot\hat M \cdot \tilde \varphi(0)
\end{eqnarray}
where matrix $\hat M$ is given by
\begin{eqnarray}\label{sol}
\hat M{=}\left(
\begin{array}{ccc}
e^{-i\frac{E_Ct}{2}}+\theta_- & -\frac{i4\Delta_m}{\xi}\sin\left(\frac{\xi t}{2}\right) & -e^{-i\frac{E_Ct}{2}}+\theta_-\\
-\frac{i4\Delta_m}{\xi_n}\sin\left(\frac{\xi t}{2}\right)  & 2\theta_+ & -\frac{i4\Delta_m}{\xi_n}\sin\left(\frac{\xi t}{2}\right)  \\
-e^{-i\frac{E_Ct}{2}}+\theta_- & -\frac{i4\Delta_m}{\xi}\sin\left(\frac{\xi t}{2}\right)  & e^{-i\frac{E_Ct}{2}}+\theta_-
\end{array}\right).\nonumber\\
\end{eqnarray}
For parametrization of the matrix $\hat M$ in Eq.(\ref{sol}) we use the shorthand notations  $\theta_{\pm}{=}\cos(\xi t/2){\pm }i(E_C/\xi)\sin(\xi t/2)$ and $\xi=\sqrt{E_C^2+8\Delta_m^2}$. In case of small driving amplitude $A{\ll} \Omega/E_C$, the Bessel function $J_m(z{\ll} 1)\approx z^m/m!$ and therefore $\Delta_m{\approx} \Delta (2AE_C/\Omega_D)^m/m!$.  
The transition probability between $\vert i \rangle$ (occupied at $t{=}{-}\infty$)  
and $\vert j\rangle$ (empty if $j{\neq} i$) states 
$P^{(m)}_{i\rightarrow j}=|\tilde \varphi_j^{(m)}(t)|^2$ for the $m$-photon resonance
is straightforwardly obtained from Eq.(\ref{solu}) and Eq.(\ref{sol}).
Assuming that either $N{-} 1$  or $N{+}1$ charge state
was occupied at  $t{=}{-}\infty$ we find that the time-depended population difference
(equivalent to the time evolution of the expectation value of $\hat S^z(t)$ operator)
is given by a slowly varying oscillating function 
\begin{eqnarray}\label{osc}
&&P^{(m)}_{1-3}=|\tilde\varphi^{(m)}_1(t)|^2-|\tilde\varphi^{(m)}_3(t)|^2\approx \\&&
\cos\left(\frac{2\Delta_m^2t}{E_C}\right)\left(1-\frac{2\Delta_m^2}{E_C^2}\right)+\frac{2\Delta_m^2}{E_C^2}\cos(E_C t).\nonumber
\end{eqnarray}
If the initial condition in Eq. (\ref{solu}) and Eq. (\ref{sol}) assumes that
the $N$-charge states is occupied  while $N{\pm}1$ states are empty, 
the oscillations
in the population difference (precession of the expectation value of $\hat S^z$)
are absent $P_{1-3}^{(m)}{=}0$. 

It is convenient to define a Fourier transform of the probability
\begin{eqnarray}\label{four}
P^{(m)}_{1-3}(\omega)=\int_{-\infty}^{+\infty} P^{(m)}_{1-3}(t)
e^{-i\omega t} dt
\end{eqnarray}
This function for the indirect $(N+1){\leftrightarrow}(N-1)$ transition
contains two Lorentzian peaks (in the presence of decoherence): 
one main peak at the frequency
$\omega{=}\Omega_R{=}2\Delta_m^2/E_C$ with a height 
$1-2(\Delta_m/E_C)^2$ and one satellite peak at $\omega{=}E_C$ with a 
height $2(\Delta_m/E_C)^2$. The Fourier transform of the 
total transition probability obtained by summation over all multi-photon processes will have a characteristic shape of a frequency comb.


\section{Summary and discussions}
The standard investigation of a Cooper-pair box model describing a charge Josephson qubit assumes projection onto a TLS near the degeneracy points
when the dimensionless gate charge $n_g$ takes the half-integer values 
$n_g{=}N{\pm} 1/2$. The degeneracy is lifted out by including a tunneling of one Cooper pair. As a result, the Landau-Zener transition with a probability controlled by the Josephson energy and Zener tunneling rate takes place.
In this paper we extended the CPB model by including an additional degeneracy point
between $N{-}1$ and $N{+}1$ Cooper pairs. The minimal model accounting for this degeneracy is formulated in terms of the three-level system. We investigated
the Landau-Zener transition associated with linear sweep of $n_g$ in the three-level model by solving the Schr\"odinger equation using Kayanuma's method. We have shown that
the LZ probabilities demonstrate a behaviour characterized by either "step" structure or "beats" pattern. We have formulated the conditions for the formation of the steps and beats in terms of the parameters of the three-level model. We introduced the mapping between the
three-level model describing the CPB and the models describing quantum dynamics of $S{=}1$ system in the presence of the single-ion anisotropy (quadrupole interaction).
Analysis of the Rabi oscillations in the periodically driven three-level system is performed in the framework of the  Rotating Wave Approximation for two important limiting cases of resonance and off-resonance drives. It is shown that if the direct transition between certain pairs of the levels is allowed by the symmetry, then the resonance Rabi oscillations are well-described by the two-level model. In that case the resonance condition assumes driving at the frequency equal to the energy offset. If, however, the direct transition between the two levels is 
forbidden by the symmetry (when the corresponding matrix element is zero), the Rabi oscillations nevertheless occur as the second order in tunneling process at
the off-resonance frequency which scales quadratically with
the Josephson energy. It is well known that for the two-level models any detuning from the resonance increases the frequency of the oscillations. The resonance condition gives a low bound for the Rabi oscillations frequency: it is equal to the amplitude of the drive.
The off-resonance Rabi oscillations in the three-level CPB Hamiltonian are predicted
to be characterized by a much smaller frequency determined by the second-order
in tunneling process. These Rabi oscillations correspond to the precession
of $S^z$ projection (the population difference between $N{+} 1$ and $N{-} 1$ states characterized by the equal odd or even parity) described by the effective $S{=}1$ Hamiltonians.\\

\section*{Acknowledgements}
We acknowledge fruitful conversations with Pertti Hakonen on 
the early stage of the project. We are grateful to Mark Dykman, Yuval Gefen, Sigmund Kohler, Heribert Lorenz, Stefan Ludwig, Valery Pokrovsky and Nikolay Sinitsyn 
for many inspiring discussion of the Landau-Zener-St\"uckelberg-Majorana physics.


\end{document}